\documentclass[11pt,twoside]{article}
\usepackage{asp2014}

\aspSuppressVolSlug
\resetcounters

\begin{document}

\title{The formation of low-mass helium white dwarfs in close binaries}
\author{Alina G. Istrate $^1$
\affil{$^1$ Argelander-Institut f\"ur Astronomie, Universit\"at Bonn, Auf
              dem H\"ugel 71, 53121 Bonn, Germany; \email{aistrate@astro.uni-bonn.de}}}

\paperauthor{Alina G. Istrate}{aistrate@astro.uni-bonn.de}{}{Universit\"at Bonn}{Astronomie}{Bonn}{}{53121}{Germany}

\begin{abstract}
Recently, a large number of low-mass ($<0.30\;M_{\odot}$) helium white dwarfs (He WDs)\footnote{ also known as extremely low-mass (ELM) white dwarfs (WDs)}have been discovered as a result of several surveys 	campaigns  as WASP, ELM, Kepler or SDSS. The far majority of them have as companion another compact object. There appears to be discrepancies between current theoretical modelling of low~-mass He WDs and a number of key observational cases indicating that some details of their formation scenario yet remain to be understood.
\end{abstract}

\section{Introduction}
    
    In recent years an increasing number of observed low-mass white dwarfs has been reported. Most of these systems have low surface gravities of $\log g\leq6.5$ and are presumed to have He cores. Single He core white dwarfs cannot be produced within a Hubble time but they are the product of binary evolution. Around 60\% of ELMs are residing in close binary systems \citep{msmm+13} such as millisecond pulsars (MSPs) or in binaries with another WD, sdB or even main sequence stars. 
 Some of these systems might be the progenitors of type Ia supernovae \citep{it+84}, underluminous .Ia supernovae \citep{bswn+07} or  Am CVn systems  \citep{khgb+14}. Because of their compact orbit they also represent perfect environments as verification sources for future gravitational wave detectors.\\
  There are a few  puzzling issues regarding the properties of recently discovered He WDs: they seem to be metal rich \citep[e.g.][]{gdkb14}, while some of them are bloated and have higher temperatures as expected for He WDs. 
  
    The existence of low-mass He WDs as companions in MSP systems has been known for a few decades \citep{kbjj+05}. In such systems, several attempts have been made to calibrate the WD cooling models on the basis of the spin-down properties of their radio MSP companion \citep{ashp96}. Assuming that the radio MSP is activated at the same time as the WD is formed, following an episode of mass transfer in a low-mass X-ray binary (LMXB), the characteristic age of the MSP \footnote{$\tau_{\rm PSR}\equiv P/(2\dot{P})$, where $P$ is the spin period and $\dot{P}$ is the period derivative} should be equivalent to the cooling age of the WD, $\tau_{\rm cool}$. However, it has been demonstrated that the characteristic age is not a good true age indicator \citep{tlk12,tau12}.
 With the discovery of the intriguing system PSR J1012+5307 \citep{nll+95}, which has $\tau_{\rm PSR}>20\;\tau_{\rm cool}$, an intense discussion started   about the WD cooling ages and MSP birthrates. Searching for an explanation for why $\tau_{\rm cool}\ll\tau_{\rm PSR}$, it was realized in the subsequent years that very low-mass proto-WDs ($<0.20\;M_{\odot}$) avoid hydrogen shell flashes whereby a relatively thick ($\sim 10^{-2} M_{\odot}$) hydrogen envelope remains intact, causing residual hydrogen shell burning to continue on a Gyr timescale. Hence, these WDs remain hot for a very long time and result in strongly underestimated cooling ages when calibrated to the conventional models available at the time. Moreover, after the LMXB phase, the donor star initially evolves through a proto-He WD phase before entering the cooling track, on a timescale depending on the WD mass (as shown in \citet{itla14}).
 
 Traditionally, WD cooling models have been constructed by artificially removing mass from the envelope of a low-mass giant star until it evolves
across the Hertzsprung-Russell  diagram and settles on the WD cooling branch \citep[e.g.][]{dsbh98,asb01}. This method often results in very small hydrogen envelope masses of $<10^{-3}\;M_{\odot}$ \citep{dsbh98}. However, it was demonstrated by \citet{seg00} that for very low-mass
He~WDs a hydrogen rich layer ($X_{\rm surf}=0.3-0.5$) of typically $0.01-0.02\;M_{\odot}$ is left behind after detachment from the Roche lobe.

A key issue for the remaining amount of hydrogen left in the envelope of a He~WD, and thus significant for its cooling timescale, is the  
question of hydrogen shell flashes \citep[see][for a detailed discussion]{ndm04}. It has been known for many years that a thermal runaway
flash may develop due to unstable CNO burning when the proto-WD evolves toward the cooling track \citep{it86,kw67,web75}. 
During these flashes the luminosity becomes very high whereby the hydrogen content is decreased via burning, 
and for strong flashes even in combination with mass loss via an additional Roche-lobe overflow \citep{it86,seg00,prp02,ndm04}. 
Whether or not hydrogen shell flashes occur depends on the WD mass, its chemical composition and the treatment of diffusion
(mixing) of the chemical elements \citep[e.g.][]{dsbh98,seg00,asb01,ndm04,amc13}.

The minimum mass ($M_{\rm flash}\simeq 0.2\;M_{\odot}$) for which these flashes occur is very important to determine. It has  previously been suggested \citep[e.g.][]{kbjj+05} that $M_{\rm flash}$  leads to a dichotomy for the subsequent WD cooling such that WDs with a mass below this limit remain hot on a Gyr timescale as a result of continued residual hydrogen shell burning, whereas WDs with a mass above this limit cool relatively fast as a result of the shell flashes eroding away the hydrogen envelope.
However, \citet{itla14} have demonstrated from theoretical modelling of LMXB evolution that this is not the case. These results will be further elaborated in the rest of this paper.

\section{Numerical methods and results}
Numerical calculations with a detailed stellar evolution code were used to follow the evolution of LMXBs systems as described in \cite{itl14}. The grid of models consists of systems with different initial donor mass, neutron star mass, orbital separation as well as different values of the magnetic braking index.
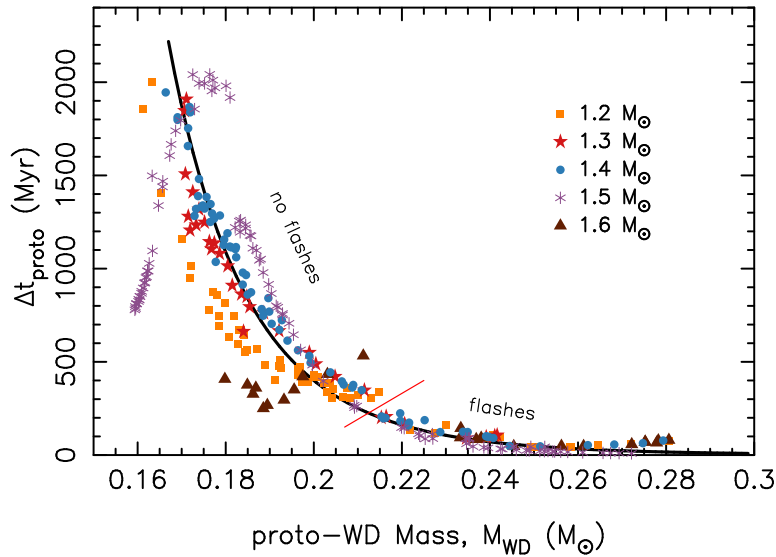
\articlefigure[width=.55\textwidth, angle=-90]{mass_deltat_mixed.ps}{deltat_mixed}{The contraction timescale, $\Delta t_{\rm proto}$ of evolution from Roche-lobe detachment until settling on the WD cooling track, plotted as a function of WD mass, $M_{\rm WD}$. The initial ZAMS masses of the WD progenitors (the LMXB donor stars) are indicated with various symbols and colors. The black line is an fit to the numerical data. The red line marks $M_{\rm flash}\simeq 0.21\;M_{\odot}$ for progenitor stars $\la 1.5\;M_{\odot}$.}

The final WD mass obtained this way is in the range of  $\sim 0.15\leq M_{\rm WD}/M_{\odot} \leq 0.30$.  
 We define $\Delta t_{\rm proto}$ as the time  it takes for the proto-WD to cross the HR diagram from the end of the mass transfer until it reaches the maximum effective temperature. Figure~\ref{deltat_mixed} shows $\Delta t_{\rm proto}$ as a function of the WD mass. One can see  that $\Delta t_{\rm proto}$ depends strongly on $M_{\rm WD}$. For very low-mass WDs that avoid hydrogen flashes ($M_{\rm WD}<M_{\rm flash}$) $\Delta t_{\rm proto}$ may be as long as 2~ Gyr, result which explains the bloated WDs observed recently.
In order to understand the differences in $\Delta t_{\rm proto}$, in Fig.~\ref{hydrogen} it is plotted the total amount of hydrogen remaining in the envelope of the proto-WD at two epochs i) at the moment of the Roche-lobe detachment and, ii) for stars which undergo hydrogen flashes, at the time the star reaches its maximum value of $T_{\rm eff}$ after the last flash. At first sight, one may suggest that this figure gives evidence for hydrogen shell flashes eroding away the hydrogen. However, as shown by \citet{itla14}, only $\sim10\%$ of the residual hydrogen is burned during the flash episode(s). The far majority ($\sim70\%$) is burned while the proto-WD crosses the HR-diagram and the remaining $\sim20\%$ is finally burned on the WD cooling track after reaching the maximum $T_{\rm eff}$. The value of $\Delta t_{\rm proto}$ is indeed  related to the ratio between the amount of hydrogen left in the envelope and the rate at which this burns (i.e. the luminosity of the proto-WD, corrected for gravitational binding energy released during the contraction phase).
 Using a modified correlation between the degenerate core mass of an evolved low-mass star and its luminosity \citep{rw71} and assuming that at the end of Roche-lobe phase the amount of hydrogen left in the envelope is always  $\sim\!0.01\pm0.005\;M_{\odot}$, one can derive an analytical relation 
 \begin{equation}
  \Delta t_{\rm proto} \simeq 400 \;\;{\rm Myr}\;\;\left(\frac{0.20\;M_{\odot}}{M_{\rm WD}}\right)^7 .
\label{eq:t_proto}
\end{equation}
For this relation we have used $L\propto M_{\rm WD}^7$, from fitting to our numerical results.
It has been shown previously in the literature that $\Delta t_{\rm proto}$ can be as large as 2 Gyr for a few single models. Our calculations show for the first time its systematic dependence on the mass of the WD (for more details see \citet{itla14}).

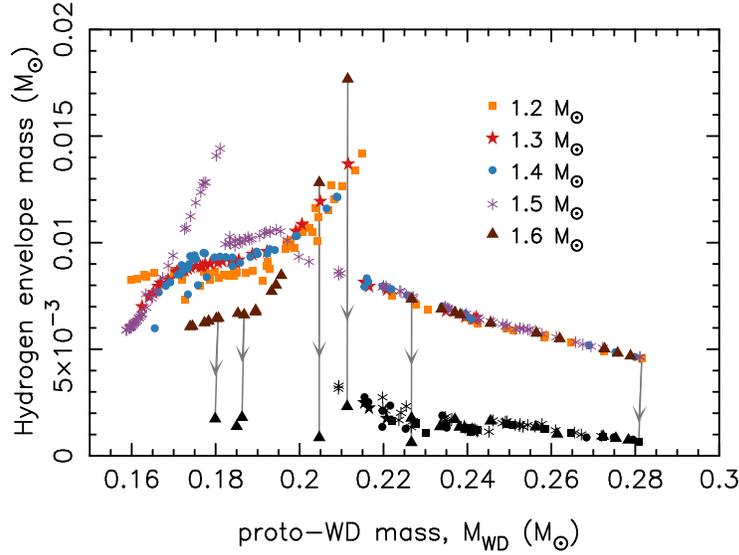
\articlefigure[width=.55\textwidth, angle=-90]{hydrogen_content_last_det.ps}{hydrogen}{Total amount of hydrogen in the envelope, $M_{\rm H-env}$ as a function of $M_{\rm WD}$ at the moment of Roche-lobe detachment for the systems plotted in Fig.~\ref{deltat_mixed}. For some of the He~WDs which experience hydrogen flashes the black symbols indicate $M_{\rm H-env}$ after settling on the WD cooling track (i.e. when they attain their max $T_{\rm eff}$). Grey arrows are examples which link the same WDs before and after the flashes (including the burning of hydrogen when crossing the HR-diagram, see text).}

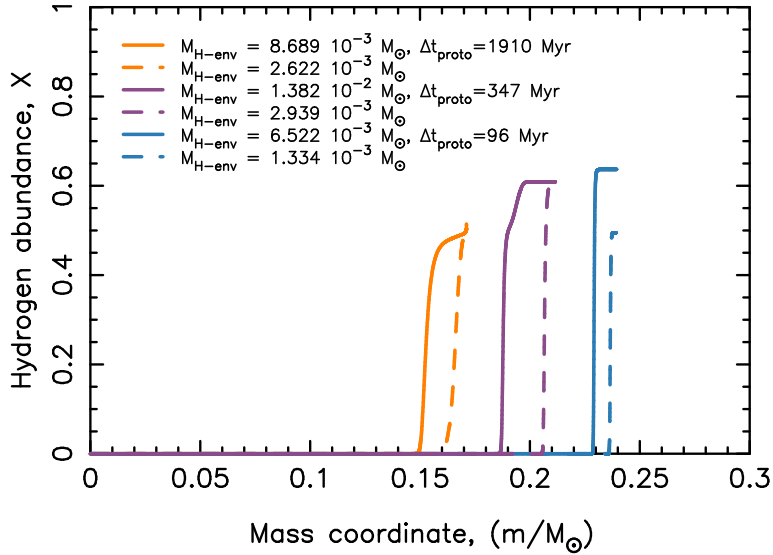
\articlefigure[width=.55\textwidth, angle=-90]{hydrogen_abundance.ps}{hydrogen_profile}{Hydrogen abundance profile of three He~WDs of mass 0.17 (no flash), 0.21 (close to the flash limit) and 0.24 $M_{\odot}$ (flash) at the end of LMXB-phase (solid line) and at  maximum $T_{\rm eff}$ (dashed line). In the legend, one can see the amount of hydrogen burned in the proto-WD phase for a system with mass below the flash limit and for two systems which experience hydrogen flashes}

The hydrogen shell flash triggered by unstable CNO burning causes a large flux of energy flowing outward.
Therefore the layers immediately above the burning shell become unstable against convection.
This causes mixing of the elements (hydrogen-rich envelope material goes down into the burning zone and 
metals are transported the other way toward the surface). This may explain the peculiar metallicity abundances recently observed in some of the low-mass He WDs, especially given that the gravitational settling timescale is relatively long in bloated envelopes.
\section{Conclusions}
Low-mass, detached proto He~WDs may spend up to 2 Gyr in the transition (contraction) phase from  Roche-lobe detachment until they reach the WD cooling track. Hence, a fair number of WDs are expected to be observed in this bloated phase, in agreement with observations.  
The timescale of this transition phase, $\Delta t_{\rm proto}$, and to some extent the minimum mass for WDs to experience hydrogen shell flashes, depend on the initial donor mass, $M_{\rm 2}$. 

In the current study, hydrogen shell flashes are found to occur for proto-WDs with $0.21\leq M_{\rm WD}/M_{\odot}\leq 0.28$ for $M_{\rm 2}\leq 1.5 M_{\odot}$. This result is in excellent agreement with the interval found by \citep{ndm04} for donors with solar metallicity and in good agreement with earlier work by \citep{dsbh98}.
We find no evidence for  $\Delta t_{\rm proto}$ to depend on the occurrence of flashes and thus question the suggested dichotomy in the thermal evolution of proto-WDs.

\acknowledgements The author is grateful for many helpful discussions with Thomas Tauris and Norbert Langer. 

\bibliography{AIstrate}  

\end{document}